\documentclass[a4paper,12pt,reqno]{amsart}
\usepackage[utf8]{inputenc}
\usepackage{amsmath}
\usepackage{amsthm,amssymb,amsfonts}
\usepackage{graphicx}
\usepackage{hyperref}
\usepackage{url}

\usepackage{geometry}
\usepackage{array,bm}
\usepackage{color}
\usepackage{verbatim}

\newtheorem{theorem}{Theorem}
\newtheorem{definition}[theorem]{Definition}
\newtheorem{corollary}[theorem]{Corollary}

\newtheorem{remark}[theorem]{Remark}

\makeatletter
\@namedef{subjclassname@2020}{\textup{2020} Mathematics Subject Classification}
\makeatother

\begin{document}
\title[Thermodynamic UR for $f$-divergence entropy production]{Thermodynamic Uncertainty Relation for $f$-divergence entropy production}

\author{Yi C. Huang} 
\address{School of Mathematical Sciences, Nanjing Normal University, Nanjing 210023, People's Republic of China}
\email{Yi.Huang.Analysis@gmail.com}
\urladdr{https://orcid.org/0000-0002-1297-7674}
\author{Xinhang Tong}
\address{School of Mathematical Sciences, Nanjing Normal University, Nanjing 210023, People's Republic of China}
\email{letterwoodtxh@gmail.com}

\date{\today} 

\subjclass[2020]{Primary 26D10. Secondary 26B12, 35A23.}  
\keywords{Thermodynamic entropy production, uncertainty relations, Kullback-Leibler divergence, $f$-divergence, Jensen inequality, convexity}

\thanks{Research of YCH is partially supported by the National NSF grant of China (no. 11801274), 
the JSPS Invitational Fellowship for Research in Japan (no. S24040), and the Open Projects from Yunnan Normal University (no. YNNUMA2403) and Soochow University (no. SDGC2418). Both authors thank in particular Professor Yoshihiko Hasegawa for his kind encouragement.}

\maketitle
\begin{abstract}
We propose an $f$-divergence extension of the Hasegawa-Nishiyama thermodynamic uncertainty relation. More precisely, we introduce the stochastic thermodynamic entropy production based on generalised  $f$-divergences and derive corresponding uncertainty relations in connection with the symmetry entropy.
\end{abstract}

\section{Introduction}
Recently, Hasegawa and Nishiyama \cite{hasegawa2025thermodynamicentropicuncertaintyrelation} used the Kullback-Leibler divergence based entropy production to answer positively the following question: 
$$\text{Does an entropic uncertainty relation hold in stochastic thermodynamics?}$$ 
To state their uncertainty relations we need recall some necessary notions.

\begin{definition}\label{defabsolute}
Let $\Phi$ be a real-valued random variable, with $|\Phi|$ being its absolute value. The probability distribution of $|\Phi|$ is given by
\begin{equation}
P(|\Phi|=\phi)=
\begin{cases}
P(\Phi = \phi) + P(\Phi = -\phi) ,& \phi > 0 ,\\
P(\Phi = 0) ,& \phi = 0 .
\end{cases}
\end{equation}
\end{definition}

Following Hasegawa and Nishiyama \cite{hasegawa2025thermodynamicentropicuncertaintyrelation}, we also introduce
\begin{definition}
Consider a real-valued observable $\Phi$, the entropy difference between the distribution $P(\Phi)$ and its absolute value distribution $P(|\Phi|)$ is defined as 
\begin{equation}
\Lambda[P(\Phi)]:=H[P(\Phi)]-H[P(|\Phi|)],
\end{equation}
where  
\begin{equation}
H[P(X)]=-\sum_{x}P(X=x)\ln P(X=x), 
\end{equation}
denotes the Shannon entropy of $P(X)$. Here we refer to $\Lambda[P(\Phi)]$ as \textbf{Symmetry Entropy} since it  measures in a sense the symmetry of the value distribution of the real-valued $\Phi$ around 0. In particular, if $\Phi\ge 0$ or $\Phi\le 0$, then $\Lambda[P(\Phi)]\equiv 0$.
\end{definition}

\begin{remark}
For continuous random variables, the Shannon entropy is defined by
\begin{equation}
H[P(\Phi)] = - \int_{-\infty}^{\infty}  P(\phi) \ln P(\phi)\,d\phi.
\end{equation}
We assume $P(\phi)$ is differentiable.
The probability density $P(|\Phi|)$ is defined by
\begin{equation}
P(|\Phi|=\phi)=P(\Phi=\phi)+P(\Phi=-\phi).
\end{equation}
We do not consider the term $P(\Phi=0)$, because such measure is necessarily 0 for the smooth probability density. 
The symmetry entropy $\Lambda[P(\Phi)]$ is then defined similarly.
\end{remark}

Assuming the local detailed balance, it is known that the entropy production under the steady-state condition can be naturally defined by \cite{hasegawa2025thermodynamicentropicuncertaintyrelation, PhysRevLett.95.040602, Seifert2008}:
\begin{equation} \label{e:prod}
\Sigma_{\text{KL}}=D_{\text{KL}}[\mathcal{P}(\Gamma)\parallel \mathcal{P}(\Gamma^{\dagger})],
\end{equation}
where 
\begin{equation}
D_{\text{KL}}[P\parallel Q]=\sum_{x}P(x)\ln\frac{P(x)}{Q(x)},
\end{equation}
is the Kullback-Leibler divergence of two (discrete) probability distributions $P$ and $Q$, $\Gamma$ is a stochastic trajectory of the process to be observed ($\Gamma^{\dagger}$ being its time reversal), 
and $\mathcal{P}(\Gamma)$ and $\mathcal{P}(\Gamma^{\dagger})$ denote the probability of measuring $\Gamma$ and $\Gamma^{\dagger}$, respectively.

The following thermodynamic entropic uncertainty relation is obtained in \cite{hasegawa2025thermodynamicentropicuncertaintyrelation}.

\begin{theorem} [Hasegawa and Nishiyama, 2025]
Let $\Sigma_{\text{KL}}$ be the entropy production as defined in \eqref{e:prod} under the steady-state condition.
Consider a real-valued observable $\Phi(\Gamma)$, which is assumed to be anti-symmetric with respect to the time reversal:
\begin{equation}\label{antisy}
\Phi(\Gamma)=-\Phi(\Gamma^{\dagger}),
\end{equation}
and let $P(\Phi)$ be the probability distribution of $\Phi$. Then we have
\begin{equation}
\Sigma_{\text{KL}} \ge[1-P(\Phi =0)]\ln 2-\Lambda[P(\Phi)]\ge 0.
\end{equation}
\end{theorem}
\begin{remark}
We refer to the first inequality as the uncertainty relation between $\Sigma_{\text{KL}}$ and $\Lambda[P(\Phi)]$,
since it indicates that their sum in the case $P(\Phi =0)=0$ is no less than $\ln 2$. 
We also point out that the quantity $[1-P(\Phi =0)]\ln 2-\Lambda[P(\Phi)]$ is nothing but the Jensen-Shannon divergence for $\Phi$ and $-\Phi$, see \cite[Appendix B]{hasegawa2025thermodynamicentropicuncertaintyrelation}.
\end{remark}
In this note, we use the general $f$-divergences, instead of the particular Kullback-Leibler divergence, to demonstrate  that there is an entropic uncertainty relation between generalised entropy production and the symmetry entropy.

\begin{definition}[e.g., \cite{Sason2016}]
Let $f:(0,\infty)\to \mathbb{R}$ be a convex function with $f(1)=0$. Let $P$ and $Q$ be two probability distributions on a measurable space $(\mathcal{X}, \mathcal{F})$. 
If $P\ll Q$ then the $f$-divergence (for $P$ over $Q$) is defined as
\begin{equation}
D_{f}[P\parallel Q]=\mathbb{E}_{Q}\left[f\left(\frac{dP}{dQ}\right)\right],
\end{equation}
where $\frac{dP}{dQ}$ is the Radon-Nikodym derivative and $f(0):=f(0+)$.
\end{definition}

We define the \textbf{Generalised Entropy Production} based on $f$-divergence as:
\begin{equation}
\Sigma_{f}=D_{f}[\mathcal{P}(\Gamma)\parallel \mathcal{P}(\Gamma^{\dagger})].
\end{equation}
Hence the Kullback-Leibler divergence $D_{\text{KL}}$ corresponds to $D_{f}$ for $f(x)=x\ln x $.

\begin{remark}
For the discrete case, with $P(x)$ and $Q(x)$ being the corresponding probability distributions, we can write out the expectation as the following sum
\begin{equation}
D_{f}[P\parallel Q]=\sum_{x\in\mathcal X}Q(x)f\left(\frac{P(x)}{Q(x)}\right).
\end{equation}
\end{remark}

Our main result is stated as follows.

\begin{theorem}\label{mainth}
Consider a real-valued observable $\Phi(\Gamma)$, which is anti-symmetric in the sense of \eqref{antisy}. If $f(x)+\ln (1+x)$ is convex, then the following inequality holds:
\begin{equation}
\Sigma_{f} \ge[1-P(\Phi =0)]\ln 2-\Lambda[P(\Phi)].
\end{equation}
\end{theorem}

In particular, for $f(x)=x\ln x$, we recover the result of Hasegawa and Nishiyama.

\begin{corollary}
Let $$f_{1}=(x-1)^{2}\quad \text{ and }\quad f_{2}=x\ln\frac{2x}{x+1}+\ln\frac{2}{x+1},$$ 
and the associated $f$-divergences become the $\chi^{2}$-divergence and Jensen-Shannon divergence. 
Since $f_{1}+\ln(1+x)$ and $f_{2}+\ln(1+x)$ are both convex, we get immediately from Theorem \ref{mainth} the following thermodynamic uncertainty relations 
\begin{equation}
\Sigma_{\chi^{2}} \ge[1-P(\Phi =0)]\ln 2-\Lambda[P(\Phi)]
\end{equation}
and
\begin{equation}\label{eqjs}
\Sigma_{JS} \ge[1-P(\Phi =0)]\ln 2-\Lambda[P(\Phi)].
\end{equation}
\end{corollary}

We refer to \cite{OZAWA2017998} for Schr\"odinger-Robertson uncertainty relations in an algebraic framework.
We refer to \cite{Seifert_2012, VANDENBROECK20156, PhysRevLett.114.158101, PhysRevLett.116.120601, Hasegawa2023, Nishiyama_2024} and the references in \cite{hasegawa2025thermodynamicentropicuncertaintyrelation} 
for other uncertainty relations in stochastic thermodynamics.

\section{Proof of Theorem \ref{mainth}}

The Shannon entropy of $\Phi$ can be rewritten as
\begin{equation}
\begin{aligned}
H[P(\Phi)]&=-\sum_{\phi}P(\phi)\ln P(\phi)\\
&=-\sum_{\phi>0}P(\phi)\ln P(\phi)-\sum_{\phi<0}P(\phi)\ln P(\phi)-P(0)\ln P(0)\\
&=-\sum_{\phi>0}[P(\phi)\ln P(\phi)+P(-\phi)\ln P(-\phi)]-P(0)\ln P(0),
\end{aligned}
\end{equation}
where we abbreviated $P(\phi)=P(\Phi=\phi)$ for notational convenience.
With the Definition \ref{defabsolute}, the Shannon entropy of $|\Phi|$ is then given by
\begin{equation}
\begin{aligned}
H[P(|\Phi|)]&=-\sum_{\phi\ge 0}P(|\Phi|=\phi)\ln P(|\Phi|=\phi)\\
&=-P(0)\ln P(0)-\sum_{\phi>0}[P(\phi)+P(-\phi)]\ln [P(\phi)+P(-\phi)].
\end{aligned}
\end{equation}
This way, we can rewrite furthermore the symmetry entropy as
\begin{equation}\label{eqsymmetry}
\begin{aligned}
\Lambda[P(\Phi)]&=H[P(\Phi)]-H[P(|\Phi|)]\\
&=-\sum_{\phi>0}P(\phi)\ln \frac{P(\phi)}{P(\phi)+P(-\phi)}-\sum_{\phi>0}P(-\phi)\ln \frac{P(-\phi)}{P(\phi)+P(-\phi)},
\end{aligned}
\end{equation}
and the $f$-divergence as 
\begin{equation}\label{eqdf}
\begin{aligned}
D_{f}[P(\Phi)\parallel P(-\Phi)]&=\sum_{\phi}P(-\phi)f\left(\frac{P(\phi)}{P(-\phi)}\right)\\
&=\sum_{\phi>0}P(-\phi)f\left(\frac{P(\phi)}{P(-\phi)}\right)+\sum_{\phi>0}P(\phi)f\left(\frac{P(-\phi)}{P(\phi)}\right).
\end{aligned}
\end{equation}
By adding equation \eqref{eqsymmetry} and equation \eqref{eqdf}, we get
\begin{equation}\label{e:key}
\begin{aligned}
& D_{f}[P(\Phi)\parallel P(-\Phi)]+\Lambda[P(\Phi)]\\
&\quad=\sum_{\phi>0}P(-\phi)\left[f\left(\frac{P(\phi)}{P(-\phi)}\right)-\ln \frac{P(-\phi)}{P(\phi)+P(-\phi)}\right]\\
&\qquad+\sum_{\phi>0}P(\phi)\left[f\left(\frac{P(-\phi)}{P(\phi)}\right)-\ln \frac{P(\phi)}{P(\phi)+P(-\phi)}\right]\\
&\quad=\sum_{\phi}P(\phi)\left[f\left(\frac{P(-\phi)}{P(\phi)}\right)-\ln \frac{P(\phi)}{P(\phi)+P(-\phi)}\right]-P(0)\ln 2\\
&\quad=\sum_{\phi}P(\phi)\left[f\left(\frac{P(-\phi)}{P(\phi)}\right)+\ln \left(1+\frac{P(-\phi)}{P(\phi)}\right)\right]-P(0)\ln 2\\
&\quad\ge \ln 2-P(0)\ln 2,
\end{aligned}
\end{equation}
where the last inequality involves the convexity of $f(x)+\ln(1+x)$ since by Jensen
\begin{equation}
\begin{aligned}
&\sum_{\phi}P(\phi)\left[f\left(\frac{P(-\phi)}{P(\phi)}\right)+\ln\left(1+\frac{P(-\phi)}{P(\phi)}\right)\right]\\
&\quad\ge f\left(\sum_{\phi}P(\phi)\frac{P(-\phi)}{P(\phi)}\right)+\ln\left(1+\sum_{\phi}P(\phi)\frac{P(-\phi)}{P(\phi)}\right)\\
&\quad=f(1)+\ln 2=\ln 2.
\end{aligned}
\end{equation}

Thanks to the Data Processing Inequality (see for example \cite{PolyanskiyWu2025}) for general $f$-divergences, 
while considering a transformation that maps the original random variables $X$ and $Y$ to new random variables $\widetilde{X}$ and $\widetilde{Y}$, we have
\begin{equation}
D_{f}[P(X)\parallel P(Y)]\ge D_{f}[P(\widetilde{X})\parallel P(\widetilde{Y})].
\end{equation}
Applying this to the anti-symmetric observable $\Phi$, we get
\begin{equation}
\begin{aligned}
\Sigma_{f}&=D_{f}[\mathcal{P}(\Gamma)\parallel \mathcal{P}(\Gamma^{\dagger})]\\
&\ge D_{f}[{P}(\Phi(\Gamma))\parallel {P}(\Phi(\Gamma^{\dagger}))]\\
&= D_{f}[P(\Phi)\parallel P(-\Phi)]\\
&\ge [1-P(\Phi =0)]\ln 2-\Lambda[P(\Phi)],
\end{aligned}
\end{equation}
where in the last line we used the key estimate \eqref{e:key}.

Next, we consider the continuous case. We have the following identities
\begin{equation}
\begin{aligned}
\Lambda[P(\Phi)]=&- \int_{0}^{\infty} P(\phi) \ln \frac{P(\phi)}{P(\phi)+P(-\phi)}d\phi\\
&\quad- \int_{0}^{\infty}P(-\phi) \ln \frac{P(-\phi)}{P(\phi)+P(-\phi)}d\phi
\end{aligned}
\end{equation}
and
\begin{equation}
\begin{aligned}
D_{f}[P(\Phi)\parallel P(-\Phi)]&=\int_{0}^{\infty} P(\phi) f\left(\frac{P(-\phi)}{P(\phi)}\right)\,d\phi\\
&\quad+\int_{0}^{\infty} P(-\phi) f\left(\frac{P(\phi)}{P(-\phi)}\right)\,d\phi.
\end{aligned}
\end{equation}
Thus we get 
\begin{equation}
\begin{aligned}
&\quad D_{f}[P(\Phi)\parallel P(-\Phi)]+\Lambda[P(\Phi)]\\
&=\int_{0}^{\infty} P(\phi) \left[f\left(\frac{P(-\phi)}{P(\phi)}\right)-\ln \frac{P(\phi)}{P(\phi)+P(-\phi)}\right]\\
&\qquad +\int_{0}^{\infty} P(-\phi) \left[f\left(\frac{P(\phi)}{P(-\phi)}\right)-\ln \frac{P(-\phi)}{P(\phi)+P(-\phi)}\right] d\phi\\
&=\int_{-\infty}^{\infty} P(\phi) \left[f\left(\frac{P(-\phi)}{P(\phi)}\right)+\ln \left(1+\frac{P(-\phi)}{P(\phi)}\right)\right]d\phi\\
&\ge f\left(\int_{-\infty}^{\infty}P(\phi)\frac{P(-\phi)}{P(\phi)}\,d\phi\right)+\ln\left(1+\int_{-\infty}^{\infty}P(\phi)\frac{P(-\phi)}{P(\phi)}\,d\phi\right)\\
&=f(1)+\ln 2=\ln 2.
\end{aligned}
\end{equation}
By the same arguments as in the discrete case, we obtain the following inequality
\begin{equation}
\Sigma_{f}\ge \ln 2-\Lambda[P(\Phi)].
\end{equation}

Therefore, we finish the proof of Theorem \ref{mainth}.

\bigskip

\section*{\textbf{Compliance with ethical standards}}

\bigskip

\textbf{Conflict of interest} The authors have no known competing financial interests or personal relationships that could have appeared to influence this reported work.

\bigskip

\textbf{Availability of data and material} Not applicable.

\bigskip

\bibliographystyle{alpha}
\bibliography{hasegawa2025}

\begin{thebibliography}{GHPE16}

\bibitem[BS15]{PhysRevLett.114.158101}
Andre~C. Barato and Udo Seifert.
\newblock Thermodynamic uncertainty relation for biomolecular processes.
\newblock {\em Phys. Rev. Lett.}, 114:158101, Apr 2015.

\bibitem[GHPE16]{PhysRevLett.116.120601}
Todd~R. Gingrich, Jordan~M. Horowitz, Nikolay Perunov, and Jeremy~L. England.
\newblock Dissipation bounds all steady-state current fluctuations.
\newblock {\em Phys. Rev. Lett.}, 116:120601, Mar 2016.

\bibitem[Has23]{Hasegawa2023}
Yoshihiko Hasegawa.
\newblock Unifying speed limit, thermodynamic uncertainty relation and
  heisenberg principle via bulk-boundary correspondence.
\newblock {\em Nature Communications}, 14:2828, Mar 2023.

\bibitem[HN25]{hasegawa2025thermodynamicentropicuncertaintyrelation}
Yoshihiko Hasegawa and Tomohiro Nishiyama.
\newblock Thermodynamic entropic uncertainty relation.
\newblock {\em arXiv:2502.06174}, 2025.

\bibitem[NH24]{Nishiyama_2024}
Tomohiro Nishiyama and Yoshihiko Hasegawa.
\newblock Tradeoff relations in open quantum dynamics via {R}obertson,
  {M}accone-{P}ati, and {R}obertson-{S}chr{\"o}dinger uncertainty relations.
\newblock {\em Journal of Physics A: Mathematical and Theoretical},
  57(41):415301, sep 2024.

\bibitem[OY17]{OZAWA2017998}
Tohru Ozawa and Kazuya Yuasa.
\newblock Uncertainty relations in the framework of equalities.
\newblock {\em Journal of Mathematical Analysis and Applications},
  445(1):998--1012, 2017.

\bibitem[PW25]{PolyanskiyWu2025}
Yury Polyanskiy and Yihong Wu.
\newblock {\em Information Theory: From Coding to Learning}.
\newblock Cambridge University Press, 2025.

\bibitem[Sei05]{PhysRevLett.95.040602}
Udo Seifert.
\newblock Entropy production along a stochastic trajectory and an integral
  fluctuation theorem.
\newblock {\em Phys. Rev. Lett.}, 95:040602, Jul 2005.

\bibitem[Sei08]{Seifert2008}
Udo Seifert.
\newblock Stochastic thermodynamics: principles and perspectives.
\newblock {\em The European Physical Journal B}, 64:423--431, Aug 2008.

\bibitem[Sei12]{Seifert_2012}
Udo Seifert.
\newblock Stochastic thermodynamics, fluctuation theorems and molecular
  machines.
\newblock {\em Reports on Progress in Physics}, 75(12):126001, nov 2012.

\bibitem[SV16]{Sason2016}
Igal Sason and Sergio Verdu.
\newblock $f$ -divergence inequalities.
\newblock {\em IEEE Transactions on Information Theory}, 62(11):5973--6006,
  November 2016.

\bibitem[VE15]{VANDENBROECK20156}
Christian {Van den Broeck} and Massimiliano Esposito.
\newblock Ensemble and trajectory thermodynamics: A brief introduction.
\newblock {\em Physica A: Statistical Mechanics and its Applications},
  418:6--16, 2015.
\newblock Proceedings of the 13th International Summer School on Fundamental
  Problems in Statistical Physics.

\end{thebibliography}

\end{document}